\documentclass[a4paper,twocolumn,groupcitations]{article}
\usepackage{mwe}
\usepackage{widetext}
\usepackage[T1]{fontenc}
\usepackage[ansinew]{inputenc}
\usepackage[english]{babel}
\usepackage{amsfonts}
\usepackage{amsmath}
\usepackage{array}
\usepackage{amsthm}
\usepackage{amssymb}
\usepackage{graphicx}
\usepackage{subfigure}
\usepackage{braket}
\usepackage{eucal}
\usepackage{verbatim}
\usepackage[table]{xcolor}
\usepackage{caption}
\usepackage{cite}
\usepackage{textcomp}
\usepackage{tikz}
\usetikzlibrary{positioning,arrows}
\usetikzlibrary{decorations.pathmorphing}
\usetikzlibrary{decorations.markings}
\usetikzlibrary{calc,decorations.markings}
\usetikzlibrary{arrows,shapes}
\usetikzlibrary{matrix,arrows}
\usepackage{pgfplots}
\usepackage{xparse}
\definecolor{jade}{HTML}{00A86B}
\newcommand{\be}{\begin{eqnarray}}
\newcommand{\ee}{\end{eqnarray}}

\renewcommand{\d}{\mbox{${\rm d}$}} 

\newcommand{\gn}{G_{\rm N}}

\newcommand{\Rh}{R_{\rm H}}

\title{\bf On the mass of bootstrapped Newtonian sources}
\author{Roberto~Casadio$^{ab}$\thanks{E-mail: casadio@bo.infn.it},
$\ $
Octavian~Micu$^c$\thanks{E-mail: octavian.micu@spacescience.ro}
$\ $
and
Jonas~Mureika$^{d}$\thanks{jmureika@lmu.edu}
\\
\\
$^a${\em Dipartimento di Fisica e Astronomia, Universit\`a di Bologna}
\\
{\em via Irnerio~46, 40126 Bologna, Italy}
\\
\\
$^b${\em I.N.F.N., Sezione di Bologna, I.S.~FLAG}
\\
{\em viale B.~Pichat~6/2, 40127 Bologna, Italy}
\\
\\
{\em $^b$Institute of Space Science, Bucharest, Romania}
\\
{\em P.O. Box MG-23, RO-077125 Bucharest-Magurele, Romania}
\\
\\
{\em $^d$Department of Physics, Loyola Marymount University}
\\
{\em Los Angeles, California, USA}
}
\begin{document}
\maketitle
%
%
\begin{abstract}
We show that the bootstrapped Newtonian potential generated by a uniform and
isotropic source does not depend on the one-loop correction for the matter coupling
to gravity.
The latter however affects the relation between the proper mass and the ADM mass
and, consequently, the pressure needed to keep the configuration stable.  
\par
\null
\par
\noindent
\textit{PACS - 04.70.Dy, 04.70.-s, 04.60.-m}
\end{abstract}
\section{Introduction and motivation}
\setcounter{equation}{0}
\label{Sintro}
Black holes represent problematic predictions of general relativity, particularly in that they
feature classical curvature singularities~\cite{HE,geroch}, which further seem to make hardly any sense 
in a quantum context.
One therefore expects that a complete description of gravity will be modified by quantum 
physics.
For this reason, an extension of Newtonian gravity that contains non-linear interaction
terms was developed in Refs.~\cite{Casadio:2016zpl,Casadio:2017cdv,BootN,Casadio:2019cux},
as a toy model to describe static, spherically symmetric sources in a quantum fashion.~\footnote{The
issue of building a quantum description will be tackled elsewhere~\cite{Casadio:2020ueb, Casadio:2020mch}.}
As we shall review below, the non-linear term describing the gravitational self-interaction 
is in particular obtained by coupling the gravitational potential to the Newtonian gravitational potential
energy density~\eqref{JV}.~\footnote{The same term can also be obtained by expanding the Einstein-Hilbert
action around flat space (for the details see Appendix~B of Ref.~\cite{Casadio:2017cdv}).}
For this reason, this approach is termed {\em bootstrapped\/} Newtonian gravity.
Solutions were then found corresponding to homogeneous matter distributions of radius $R$
for which no Buchdahl limit~\cite{buchdahl} appears, but still require increasingly large pressure to counterbalance the
gravitational pull for increasing compactness.
\par
Indeed, the model naturally contains two mass parameters, one which appears in the potential
outside the source and can be identified with the Arnowitt-Deser-Misner (ADM) mass~\cite{adm},
and a second mass term $M_0$ that is simply the volume integral of the proper density (from
which the energy associated with the pressure is excluded).
Since only $M$ can be measured by studying orbits around the compact object, we shall
define the compactness in terms of $M$ as $\gn\,M/R$ like in Ref.~\cite{Casadio:2019cux}.
One then obtains a unique relation between $M_0$ and $M$.
As a further development of the model, we are here interested in analysing in more detail the effects
of the couplings introduced in Ref.~\cite{Casadio:2019cux} on these two masses.

%
%
%
%
%
%
%
%
%
\par
We recall from Ref.~\cite{Casadio:2017cdv} that a non-linear equation for the potential $V=V(r)$
describing the gravitational pull on test particles generated by a matter density $\rho=\rho(r)$
can be obtained starting from the Newtonian Lagrangian~\footnote{Since all functions only depend 
on the radial coordinate $r$, we use the notation $f'\equiv\d f/\d r$.}
\be
L_{\rm N}[V]
=
-4\,\pi
\int_0^\infty
r^{2} \,\d r
\left[
\frac{\left(V'\right)^2}{8\,\pi\,\gn}
+\rho\,V
\right]
\label{LagrNewt}
\ee
and the corresponding Poisson equation of motion
\be
r^{-2}\left(r^{2}\,V'\right)'
\equiv
\triangle V
=
4\,\pi\,\gn\,\rho
\ .
\label{EOMn}
\ee
We can then include the effects of gravitational self-interaction by noting that the Hamiltonian
\be
H_{\rm N}[V]
=
4\,\pi
\int_0^\infty
r^{2}\,\d r
\left(
-\frac{V\,\triangle V}{8\,\pi\,\gn}
+\rho\,V
\right)
\ ,
\label{NewtHam}
\ee 
computed on-shell by means of Eq.~\eqref{EOMn}, yields the total Newtonian potential energy
\be
U_{\rm N}[V]
=
2\,\pi
\int_0^\infty 
{r}^{2}\,\d {r}
\,\rho(r)\, V(r)
\nonumber
\\
=
-4\,\pi
\int_0^\infty 
{r}^{2} \,\d {r}\,
\frac{\left[V'(r) \right]^2}{8\,\pi\,\gn}
\ ,
\label{Unn}
\ee
where we assumed boundary terms vanish.
Following Refs.~\cite{Casadio:2016zpl,Casadio:2017cdv,BootN,Casadio:2019cux}, one can view
$U_{\rm N}$ as given by the volume integral of the gravitational current
\be
J_V
=
-\frac{\left[ V'(r) \right]^2}{2\,\pi\,\gn}
\ .
\label{JV}
\ee
We can also include the source term 
\be
J_\rho=-2\,V^2
\ ,
\ee
which comes from the linearisation of the volume measure around the vacuum~\cite{Casadio:2017cdv}
and can be interpreted as a gravitational one-loop correction to the matter density.
As we recalled above, in Ref.~\cite{BootN}, no Buchdahl limit~\cite{buchdahl} was found but the pressure
$p$ becomes very large for compact sources with a size $R\lesssim\Rh\equiv 2\,\gn\,M$, and one must
therefore add a corresponding potential energy $U_{\rm B}$ such that
\be
p
=
-\frac{\d U_{\rm B}}{\d \mathcal{V}} 
\ .
\label{JP}
\ee
This can be easily included by simply shifting  $\rho \to \rho+p$ to yield~\footnote{This way of including the
pressure is in analogy with the definition of the Tolman mass~\cite{tolman}.}
\begin{widetext}
\be
L[V]
&\!\!=\!\!&\!\!
-4\,\pi
\int_0^\infty
r^{2} \,\d r
\left[
\frac{\left(V'\right)^2}{8\,\pi\,\gn}
+
\left(V+q_\rho\,J_\rho\right)
\left(\rho+p\right)
q_V\,J_V\,V
\right]
\nonumber
\\
&\!\!=\!\!&
-4\,\pi
\int_0^\infty
r^{2}\,\d r
\left[
\frac{\left(1-4\,q_V\,V\right)\left(V'\right)^2}{8\,\pi\,\gn}
+V
\left(1-2\,q_\rho\,V\right)
\left(\rho+p\right) 
\right]
\ ,
\label{LagrV}
\ee
\end{widetext}
where the non-negative coefficients $q_V$ and $q_\rho$ play the role of coupling constants for the graviton currents $J_V$
and $J_\rho$~\footnote{Different values of $q_V$ can be implemented in order to obtain the approximate potential
for different motions of the test particles in general relativity.}.  
The associated effective Hamiltonian is simply given by
\be
\label{HamV}
H[V]
=
-L[V]
\ ,
\ee
and the Euler-Lagrange equation for $V$ is given by the modified Poisson equation
\be
\triangle V
\!=\!
4\pi\,\gn\,
\frac{1-4\,q_\rho\,V}{1-4\,q_V\,V}
\left(\rho+p\right)
+
\frac{2\,q_V\left(V'\right)^2}
{1-4\,q_V\,V}
\!\!
\ .
\label{EOMV}
\ee
We can therefore see that in this simplified bootstrapped picture, there appears an ``effective Newton constant'' 
\be
\tilde
G_{\rm eff}
=
\frac{1-4\,q_\rho\,V}{1-4\,q_V\,V}\,
\gn
\ ,
\ee
as well as an ``effective self-coupling''
\be
q_{\rm eff}
=
\frac{q_V}
{1-4\,q_V\,V}
\ .
\ee
It is interesting to note that both effective couplings decrease when the field $V$ is negative and large
if $q_\rho<q_V$, something one would expect, {\em e.g.}~in the asymptotic safety scenario~\cite{as}.
\par
The conservation equation that determines the pressure reads
\be
p'
=
-V'\left(\rho+p\right)
\ .
\label{eqP}
\ee
In the vacuum (where $\rho=p=0$), Eq.~\eqref{eqP} is trivially satisfied and Eq.~\eqref{EOMV}
is exactly solved by~\cite{BootN}
\be
V
=
\frac{1}{4\,q_V}
\left[
1-
\left(1+\frac{6\,q_V\,\gn\,M}{r}\right)^{2/3}
\right]
\ ,
\label{sol0}
\ee
where the integration constants were fixed in order to recover the Newtonian
behaviour at large distance,
\be
V_{\rm N}
=
-\frac{\gn \,M}{r}
\ .
\label{Vn0}
\ee
Note that we can now take the limit $q_V\to 0$ and precisely recover the Newtonian potential~\eqref{Vn0},
as one would expect by first considering this limit in Eq.~\eqref{EOMV}.
We also note that the large $r$ expansion of the solution~\eqref{sol0} reads
\be
V
\simeq
-\frac{\gn\,M}{r}
+q_V\,\frac{\gn^2\,M^2}{r^2}
\ ,
\ee
so that $q_V$ always affects the post-Newtonian order.
\par
In the following analysis, we are specifically interested in the effect of the one-loop coupling $q_\rho$
on the relation between the mass $M$ and the proper mass $M_0$ of the source (which we will introduce
shortly), hence we set $q_V=1$ and consider the range $q_\rho\ge 0$.
\section{Interior solutions}
\label{S:intsolution}
\setcounter{equation}{0}
In order to derive the interior potential, we proceed as in the previous
Refs.~\cite{Casadio:2016zpl,Casadio:2017cdv,BootN,Casadio:2019cux}, in which
the source is simply modelled as a spherically-uniform proper density distribution of matter
with radius $R$,
\be
\rho
=
\rho_0
\equiv
\frac{3\,M_0}{4\,\pi\,R^3}\, 
\Theta(R-r)
\ ,
\label{HomDens}
\ee
where $\Theta$ is the Heaviside step function and the total mass $M_0$ is defined as
\be
M_0
=
4\,\pi
\int_0^R
r^{2}\,\d r\,\rho(r)
\ .
\ee
\par
We use Eq.~\eqref{eqP} to express the pressure in terms of the potential itself
like in Ref.~\cite{Casadio:2019cux} as
\be
p
=
\rho_0
\left(e^{V_R-V}
-1\right)
\label{p}
\ee
and obtain
\be
\triangle V
=
\frac{3 \,\gn\, M_0}{R^3}
\left(\frac{1-4\,q_\rho\,V}{1-4\,V}\right)
e^{V_R - V}
+ \frac{2\left(V^\prime\right)^2}{1-4\,V}
\!\!
\ .
\label{EOMVint}
\ee
Regularity conditions in the centre are required to be met by the solutions, specifically
\be
V_{\rm in}'(0)=0
\ ,
\label{b0}
\ee
where $V_{\rm in}=V(0\le r\le R)$,
and they must also satisfy matching conditions with the exterior solution at the surface, 
\be
V_{\rm in}(R)
=
V_{\rm out}(R)
\equiv
V_R
=
\frac{1}{4}
\left[
1-\left(1+{6X}\right)^{2/3}
\right]
\label{bR}
\\
\notag
\\
V'_{\rm in}(R)
=
V'_{\rm out}(R)
\equiv 
V'_R
=
\frac{X}{R\,(1+6\,X)^{1/3}}
\ ,
~~~~~
\label{dbR}
\ee
where $V_{\rm out}=V(R\le r)$.
We also introduced the ``outer'' compactness
\be
X=\frac{\gn\,M}{R}
\ ,
\ee
where it is important to keep in mind that the ADM mass $M\ne M_0$ in general.
\subsection{Small and medium compactness}
We can approach the problem in a similar way as in Ref.~\cite{Casadio:2019cux} for the case when the radius
of the source $R$ is much larger or of the order of $\gn\,M$. 
An analytic approximation $V_{\rm s}$ for $V_{\rm in}$ can be obtained by expanding around $r=0$, and
thus the expression for the potential in~\eqref{EOMVint} can be written
\be
V_{\rm s}
\simeq
V_0
+
\frac{\gn\,M_0}{2\,R^2}
\left(\frac{1-4\,q_\rho\,V_0}{1-4\,V_0}\right)
e^{V_R-V_0}\,r^2
\ ,
\label{VsS}
\ee
where $V_0\equiv V_{\rm in}(0)<0$.  We also used the regularity condition~\eqref{b0},
which constrains all odd order terms in $r$ from the Taylor expansion about $r=0$ to vanish.
\par
\begin{figure}[t]
\centering
\includegraphics[width=8cm]{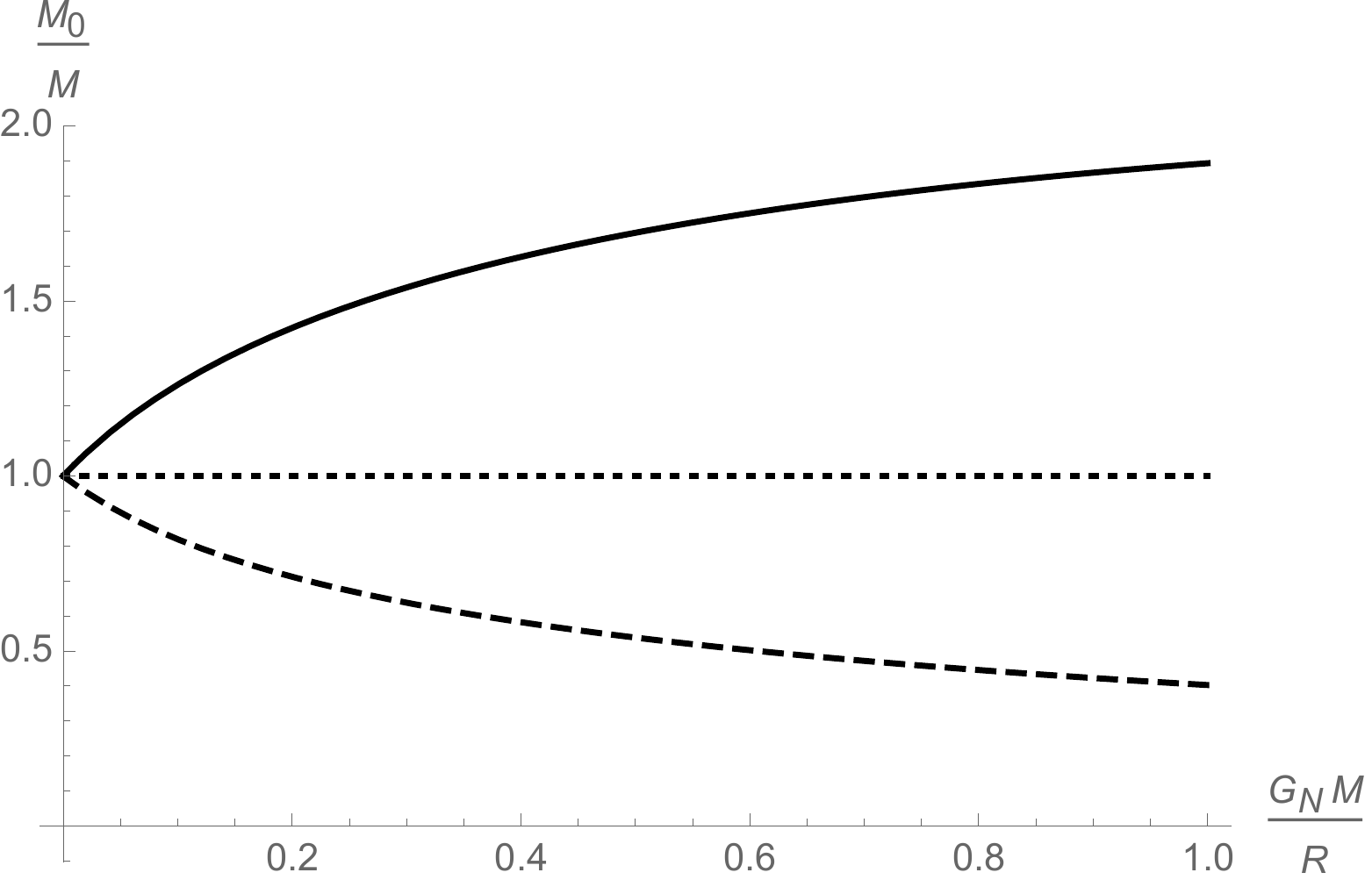}
\caption{Ratio $M_0/M$ for small and medium compactness for $q_\rho=1$ (dashed line),
and $q_\rho=0$ (solid line).
In these two cases $M_0$ is always different from $M$ (dotted line).}
\label{M0S}
\end{figure}
\begin{figure}[h]
\centering
\includegraphics[width=8cm]{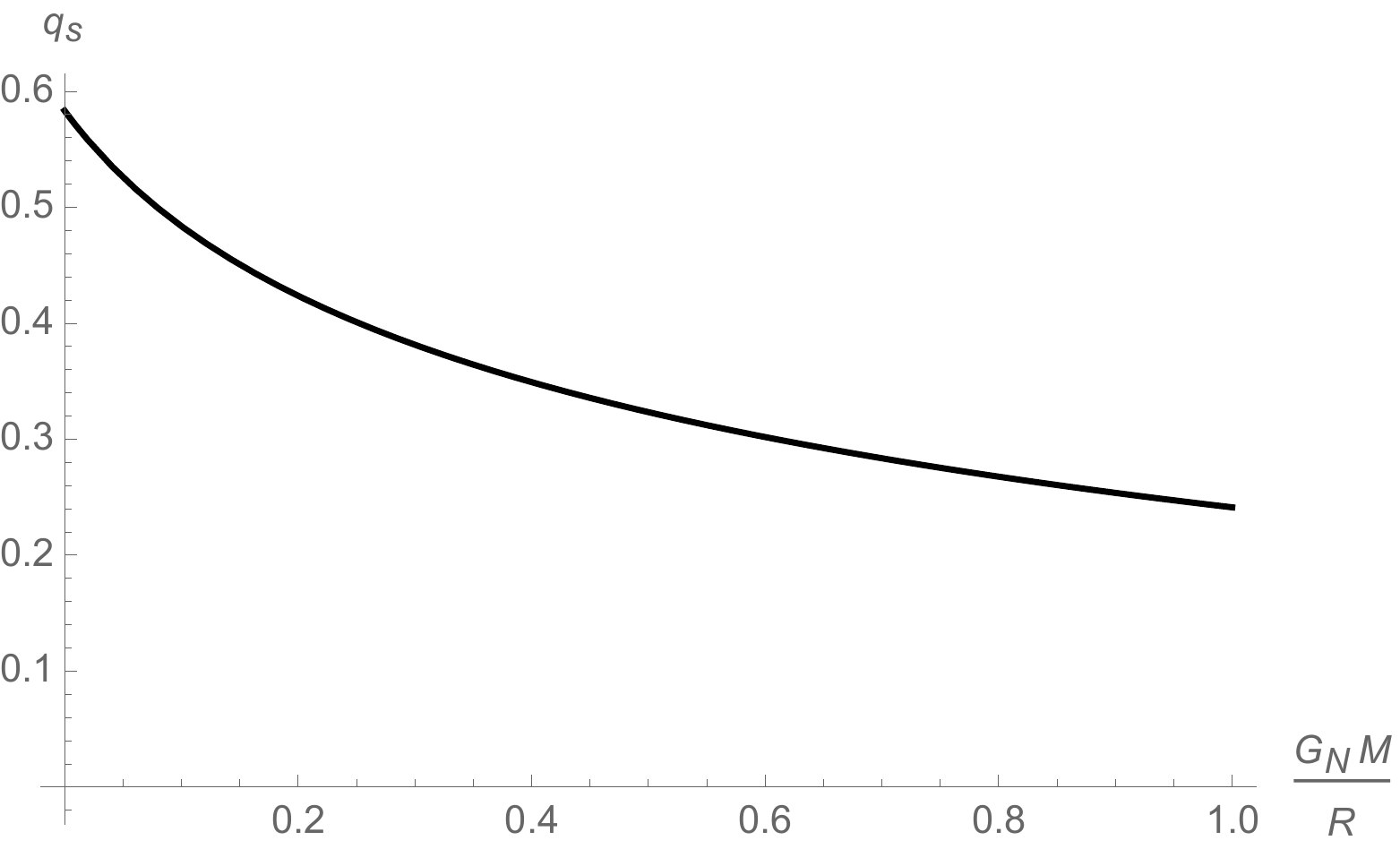}
\caption{Critical value $q_{\rm s}$ of $q_\rho$ for which  $M=M_0$ for small and medium compactness.}
\label{f:qs}
\end{figure}
\begin{figure}[h]
\centering
\includegraphics[width=8cm]{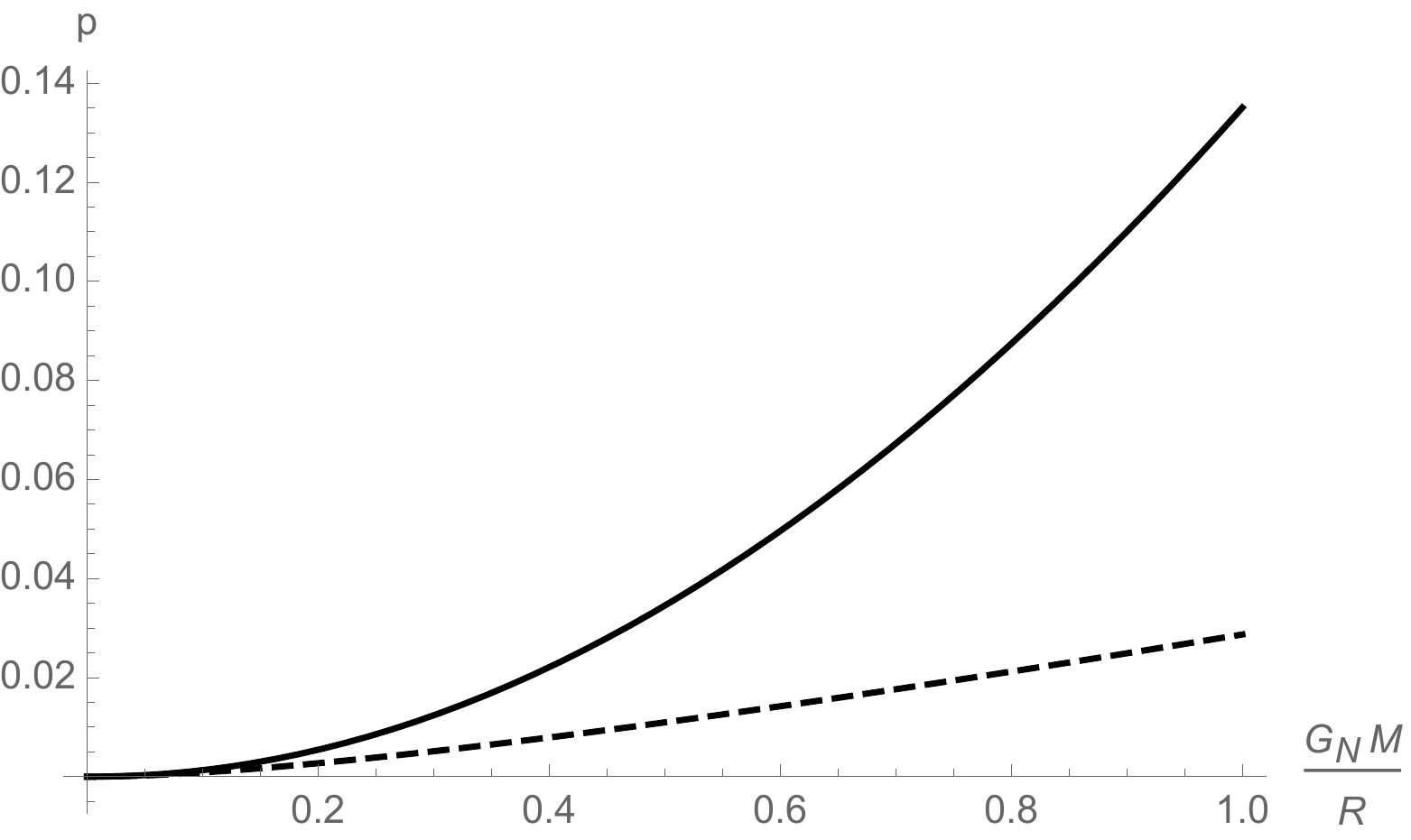}
\caption{Pressure $p$ for small and medium compactness for $q_\rho=1$ (dashed line),
and $q_\rho=0$ (solid line).}
\label{pS}
\end{figure}

After imposing the boundary conditions~\eqref{bR} and \eqref{dbR}, we find that
the potential has the same expression for any values of $q_\rho$,
\be
\label{Vins}
V_{\rm s}
\simeq
\frac{R^2\left[\left(1+6\,X\right)^{1/3}-1\right]
+2\,X\left(r^2-4\,R^2\right)}
{4\,R^2\left(1+6\,X\right)^{1/3}}
,
\label{Vintermediate_n_0}
\ee
but the relation between $M_0$ and $M$ does depend on $q_\rho$,
\be
\frac{M_0}{M}
\simeq
\frac{e^{-\frac{X}{2\left(1+6\,X\right)^{1/3}}}\left(1+8\,X\right)}
{\left(1+6\,X\right)^{2/3}
\left[
1-q_\rho
+\frac{\left(1+8\,X\right)}
{\left(1+6\,X\right)^{1/3}}
q_\rho
\right]}
\ ,
\label{M0M}
\ee
which is plotted for the two cases $q_\rho=1$, respectively $q_\rho=0$ in Fig.~\ref{M0S}. 
Different values of $q_\rho$ interpolate between these cases and a critical value of $q_\rho=q_{\rm s}$ can
be found such that $M_0=M$ (see Fig.~\ref{f:qs}),
\be
q_{\rm s}
\simeq
\frac{(1+8\,X)\,e^{-\frac{X}{2\,(1+6\,X)^{1/3}}}-(1+6\,X)^{1/3}}
{(1+6\,X)^{1/3}\left[1+8\,X-(1+6\,X)^{1/3}\right]}
\ .
\label{qs}
\ee
For $q_{\rm s}\lesssim q_\rho$ the mass $M_0<M$ as in Ref.~\cite{Casadio:2019cux},
whereas $M_0>M$ for $0\le q_\rho\lesssim q_{\rm s}$.
It is also worth noting that the pressure $p$ in Eq.~\eqref{p} grows faster with the compactness 
for $0\le q_\rho\lesssim q_{\rm s}$ than it does for $q_{\rm s}\lesssim q_\rho$ (see Fig.~\ref{pS}).
\subsection{Large compactness}
\begin{figure}[t]
\centering
\includegraphics[width=8cm]{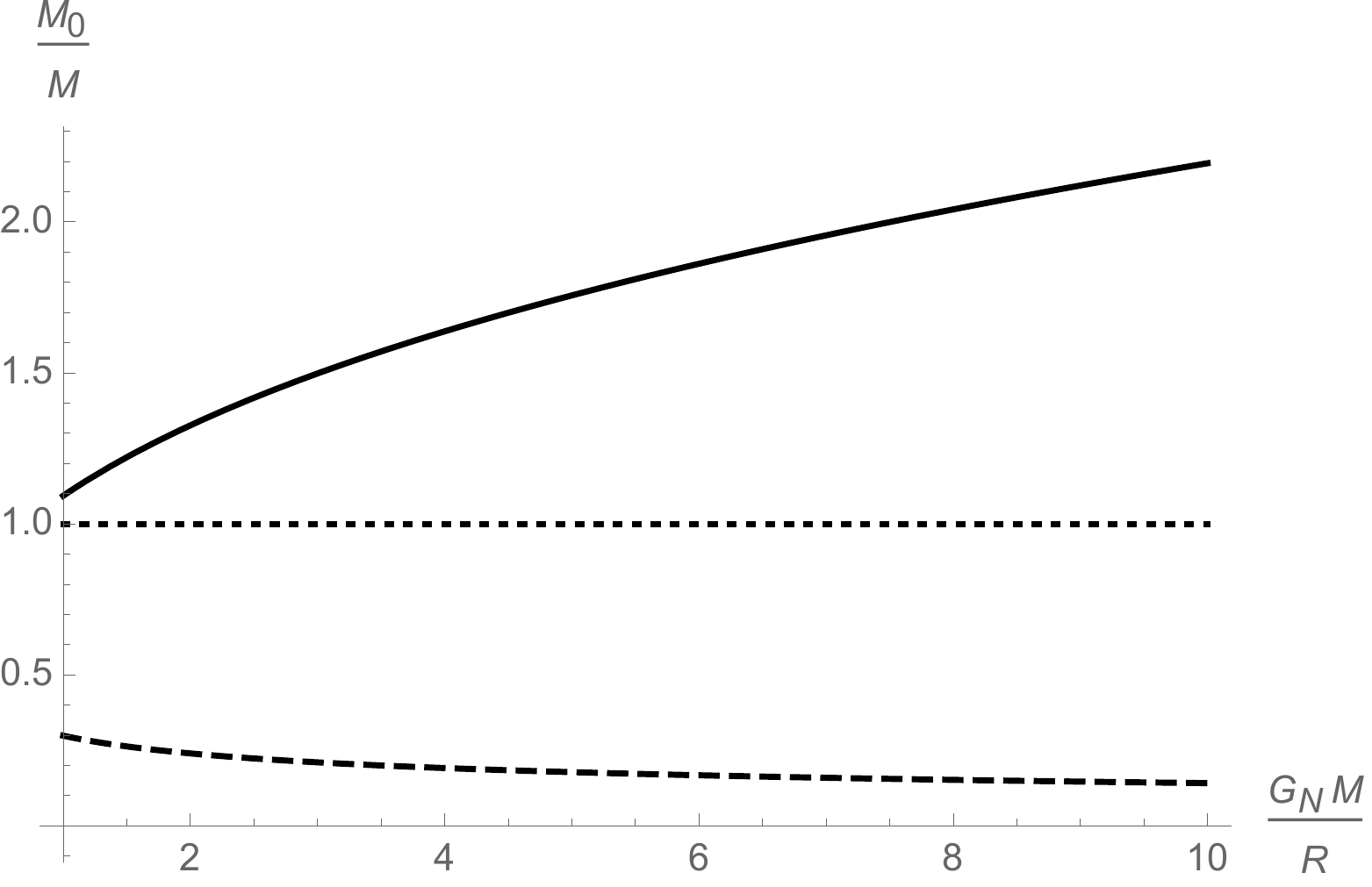}
\caption{Ratio $M_0/M$ for large compactness for $q_\rho=1$ (dashed line),
and $q_\rho=0$ (solid line).
In these two cases $M_0$ is always different from $M$ (dotted line).}
\label{M0C}
\end{figure}

\begin{figure}[h]
\centering
\includegraphics[width=8cm]{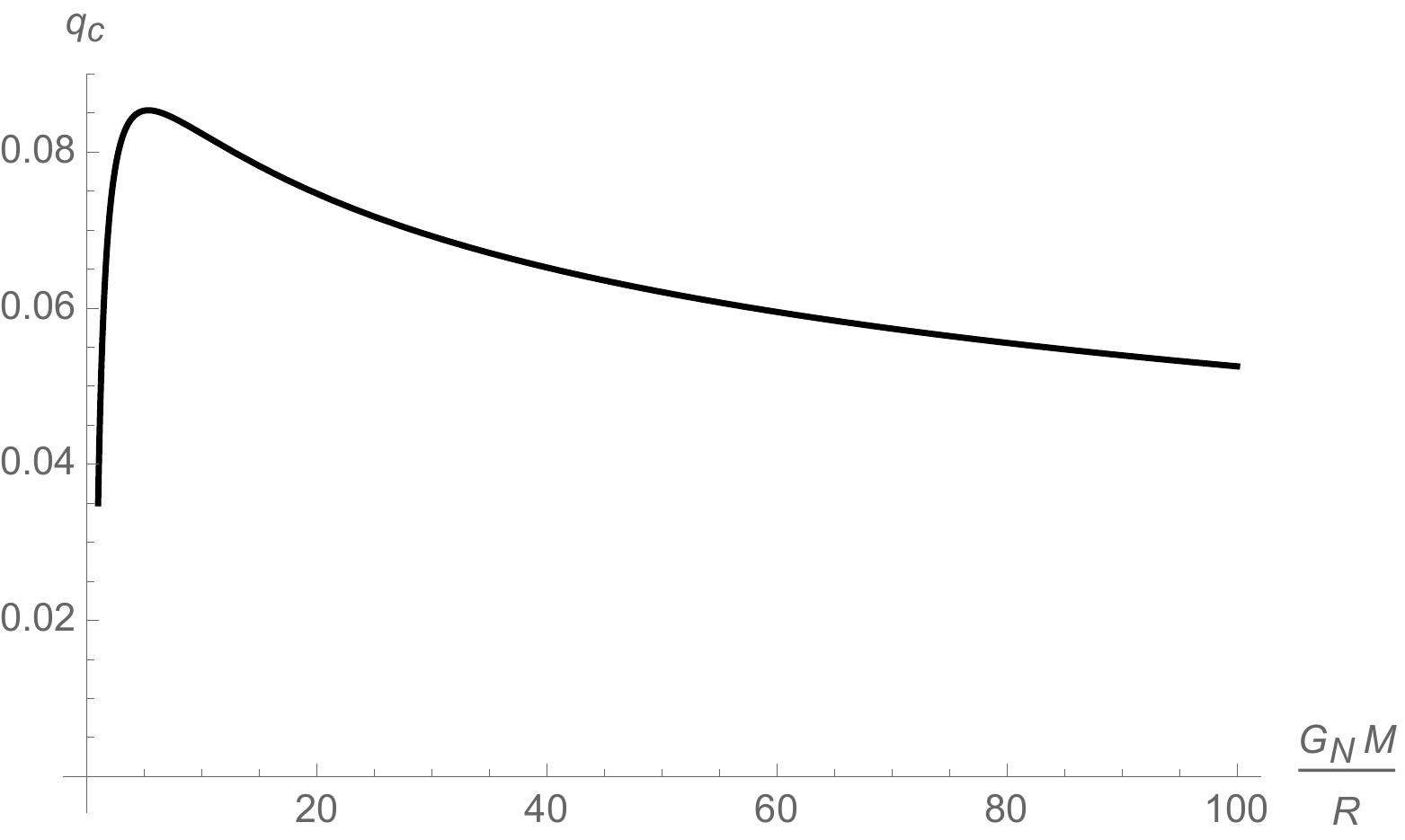}
\caption{Critical value $q_{\rm c}$ of $q_\rho$ for which  $M=M_0$ for large compactness.}
\label{f:qc}
\end{figure}

\begin{figure}[h]
\centering
\includegraphics[width=8cm]{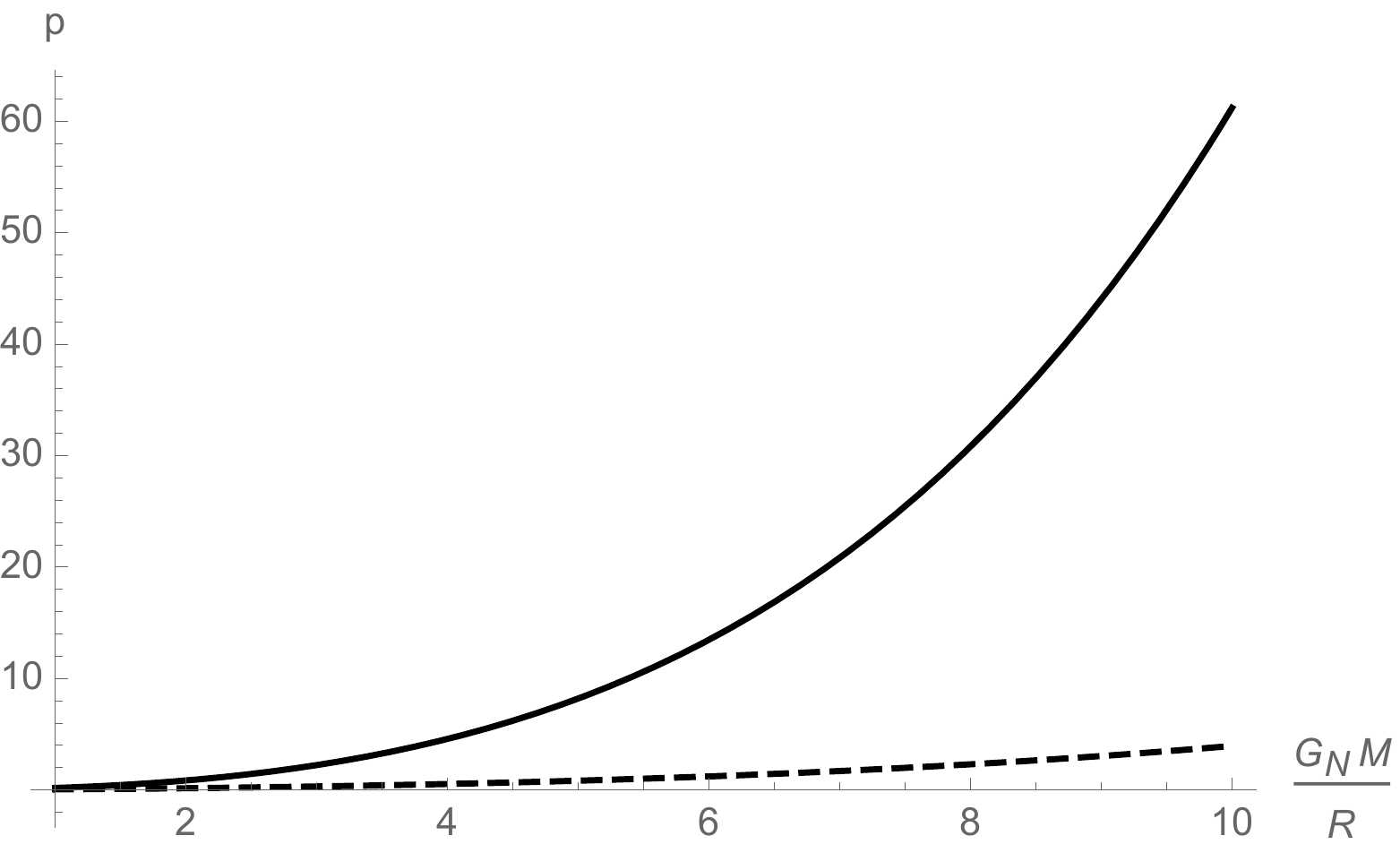}
\caption{Pressure $p$ for large compactness for $q_\rho=1$ (dashed line),
and $q_\rho=0$ (solid line).}
\label{pC}
\end{figure}

In the large compactness case, $\gn\,M\gg R$, we can employ the linear approximation~\cite{Casadio:2019cux}
\be
V_{\rm c}
\simeq
V_R
+
V_R'\,(r-R)
\ ,
\label{Vlin}
\ee
which obviously does not depend on $q_\rho$ (see Appendix~\ref{A:comparison} for more details).
The matching conditions~\eqref{bR} and \eqref{dbR} at $r=R$ are now satisfied by construction and we can hence determine
the relation between $M$ and $M_0$ by imposing the field equation~\eqref{EOMVint}, yielding
\be
\!\!\frac{M_0}{M}\!\!
\simeq\!\!
\frac{2\left(1+5\,X\right)}
{3\left(1+6X\right)^{2/3}\!\!
\left\{
1-q_\rho
\left[1
-\left(1+6X\right)^{2/3}
\right]\!
\right\}}\!\!
\ ,
\ee
which is plotted for the two cases $q_\rho=1$, respectively $q_\rho=0$, in Fig.~\ref{M0C}
and the critical value of $q_\rho=q_{\rm c}$ such that $M_0=M$,
\be
q_{\rm c}
\simeq
\frac{2\,(1+5\,X)-3\,(1+6\,X)^{2/3}}
{3\,(1+6\,X)^{2/3}\left[(1+6\,X)^{2/3}-1\right]}
\ ,
\label{qc}
\ee
is plotted in Fig.~\ref{f:qc}.
It is easy to see from Eq.~\eqref{qc} that $q_{\rm c}\sim X^{-1/3}\to 0$ for $X\to\infty$.
As with smaller values of the compactness, the mass $M_0<M$ for $q_{\rm c}\lesssim q_\rho$,
whereas $M_0>M$ for $0\le q_\rho\lesssim q_{\rm c}$, and the pressure again grows with the compactness
much faster when $M_0>M$ (see Fig.~\ref{pC}).
Finally, one should keep in mind that the linear approximation becomes rather accurate only for
values of the compactness $X\gg 1$, which explains why the ratios $M_0/M$ and
the values of $q_\rho$ for which $M_0=M$ do not match around $X=1$~\footnote{We find
that the critical couplings $q_{\rm s}$ and $q_{\rm c}$ are numerically very close for values of $X\sim 4$, and
that the masses $M_0=M_0(X)$ estimated in the two regimes are also rather close for the same
compactness.}.
\par
We have therefore shown that, not only is the outer potential insensitive to the matter coupling $q_\rho$,
but so is the interior potential (within our approximations).
Since the outer potential only depends on the ``total ADM energy'' $M$, the fact that the value of $q_\rho$
does not change it is expected.
The value of $q_\rho$, however, can affect the relation between $M_0$ and $M$ very significantly.
\section{Discussion and conclusions}
\label{S:conc}
\setcounter{equation}{0}
In this work, we focused on the effects induced by the strength of the one-loop coupling $q_\rho$
in the Lagrangian~\eqref{LagrV} on the potential $V$ generated by a static compact source of uniform density.
For this analysis, we set $q_V=1$ and values of $q_\rho$ therefore measure the 
relative strength of this contribution with respect to the gravitational self-interaction proportional
to $q_V$.
\par
The main conclusions are that a) the potential $V$ is totally insensitive to the value of $q_\rho\ge 0$
but b) the relation between the ADM mass $M$ and the proper mass $M_0$ does depend on $q_\rho$.
In particular, $M_0>M$ and the pressure necessary to keep the system in equilibrium is much larger when
$q_\rho<q_{\rm cr}$, where $q_{\rm cr}\simeq q_{\rm s}$ in Eq.~\eqref{qs} for small compactness
$\gn\,M\lesssim R$ and $q_{\rm cr}\simeq q_{\rm c}$ in Eq.~\eqref{qc} for large compactness
$\gn\,M> R$.
Since $q_{\rm cr}<1=q_V$, this case was not covered in Ref.~\cite{Casadio:2019cux}, where we
assumed $q_\rho=q_V$ and we always had $M_0<M$ accordingly.
We also remark that $q_{\rm c}\ll 1$ for very large compactness $\gn\,M\gg R$ and that it asymptotes to zero, which
makes this case somewhat less likely to play a relevant role in modelling (quantum) black holes than the case
studied in Ref.~\cite{Casadio:2019cux}.
\par
We conclude by noting that the fact the potential $V$ for static configurations does not change with $q_\rho$,
and is therefore insensitive to $M_0$, but only depends on the total mass $M$ and radius $R$ of the source
appears as a form of Birkhoff's theorem in the bootstrapped Newtonian picture.  
\section*{Acknowledgments}
R.C.~is partially supported by the INFN grant FLAG.
The work of R.C.~has also been carried out in the framework
of activities of the National Group of Mathematical Physics (GNFM, INdAM)
and COST action {\em Cantata\/}. 
O.M.~is supported by the grant Laplas~VI of the Romanian National Authority for Scientific
Research.  J.M.~thanks the Department of Physics and Astronomy at the University of Bologna for its generous 
hospitality during the initial stages of this project.
\appendix
\section{Comparison method for large compactness}
\label{A:comparison}
\setcounter{equation}{0}
Using the comparison method for non-linear differential equations, it was shown in
Ref.~\cite{Casadio:2019cux} that the linear potential~\eqref{Vlin} is a good approximation in the large
compactness regime for $q_\rho=1$, except in a (very) small region near $r=0$, where it does not satisfy
the boundary condition~\eqref{b0}.
We briefly show here that this still holds for $q_\rho\ge 0$.
\par
The comparison theorems~\cite{BVPexistence, BVPexistence1,ode} (see also Appendix~C in
Ref.~\cite{Casadio:2019cux})
ensure that the solution to Eq.~\eqref{EOMVint} must lie in between any two bounding functions,
\be
\label{vminvmax}
V_-<V_{\rm in}<V_+
\ .
\ee
which satisfy (suitably generalised) boundary conditions and are such that $E_+(r)<0$ and $E_-(r)>0$
for $0\le r\le R$, where
\begin{widetext}
\be
E_\pm
\equiv
\triangle V_\pm
-
\frac{3\,\gn\,M_0^{\pm}(M)}{R^3}\left(\frac{1-4\,q_\rho\,V_\pm}{1-4\,V_\pm}\right)e^{V_R-V_\pm}
-\frac{2\left(V_\pm'\right)^2}{1-4\,V_\pm}
\ .
\ee
\end{widetext}
For $X\equiv \gn\,M/R\gg 1$, we consider the simpler equation
\be
\psi''
=
\frac{3\,\gn\,M_0}{R^3}\,e^{V_R-\psi}
\ ,
\ee 
which is solved by
\be
\resizebox{\linewidth}{!}
\!\!\!\!\psi(r;A,B) \!\!\!\!
&=&
\!\!\!\!-A\left(B+\frac{r}{R}\right)\nonumber \\
&+&\!\!\!\!
2\ln\!\!\left[1+\frac{3\,\gn\,M_0}{2\,A^2\,R}\,e^{A\,(B+r/R)+V_R}\right]
,
\label{psiAB}
\ee
where the constants $A$, $B$ and $M_0$ are determined by
the boundary conditions~\eqref{b0}, \eqref{bR} and~\eqref{dbR}.
Regularity at $r=0$ in particular yields
\be
M_0
=
\frac{2\,A^2\,R}{3\,\gn}\,e^{-A\,B-V_R}
\ .
\label{M00}
\ee
Eq.~\eqref{dbR} for the continuity of the derivative across $r=R$ then reads
\be
A\,\tanh(A/2)
\simeq
A
=
R\,V_R'
\ ,
\label{eqA}
\ee
and he continuity Eq.~\eqref{bR} for the potential,
\be
2\,\ln\left(1+e^{R\,V_R'}\right)
-R\,V_R'\,(1+B)
=
V_R
\ ,
\ee
can be used to express $B$ in terms of $M$ and $R$.
Putting everything together, we obtain~\cite{Casadio:2019cux}
\be
\psi(r;X,R)
\simeq
\frac{1}{2}
\left(\frac{X}{\sqrt{6}}\right)^{2/3}
\left(\frac{2\,r}{R}-5\right)
\ .
\label{psi0}
\ee 
\par
\begin{figure}[t]
\centering
\includegraphics[width=8cm]{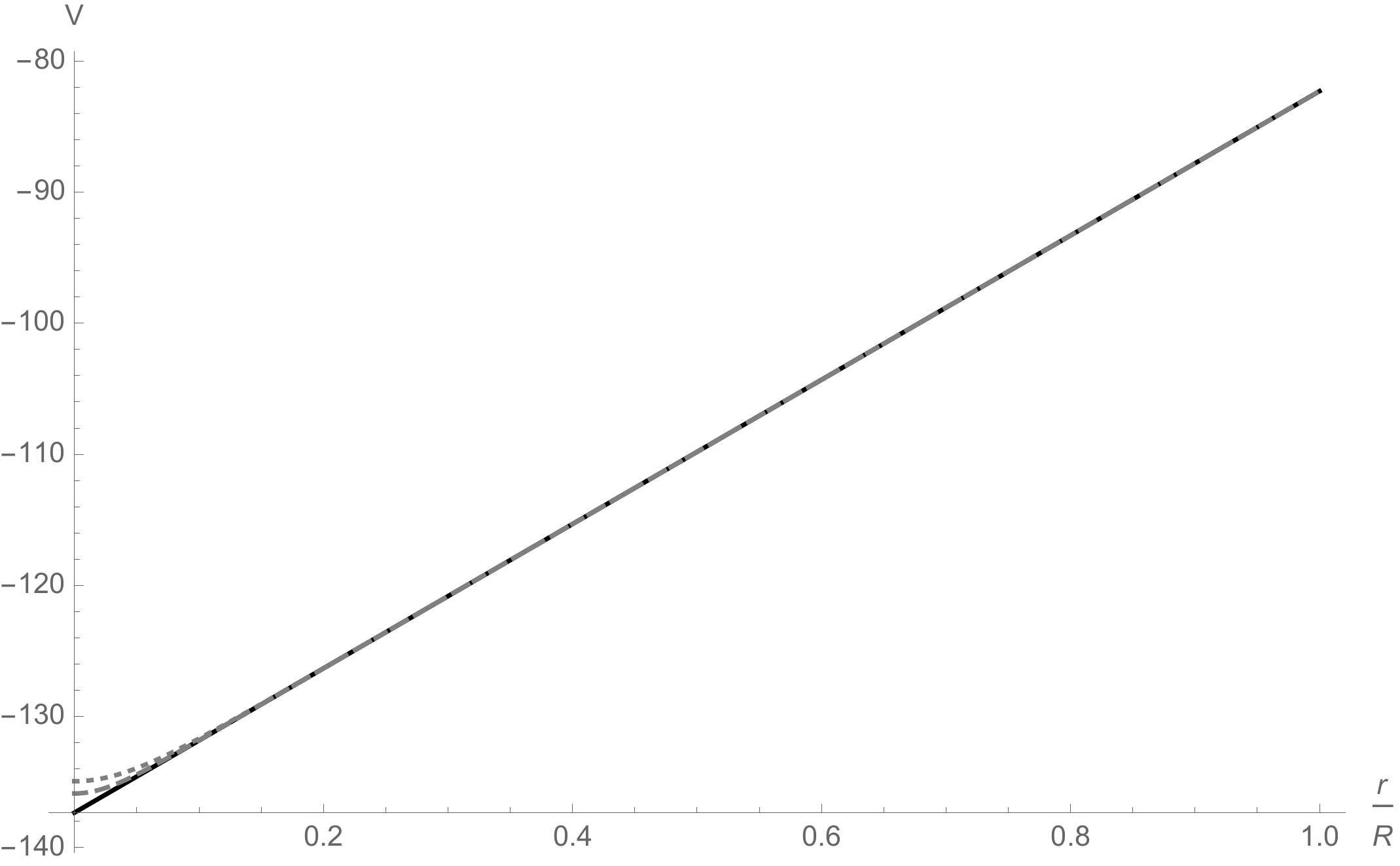}
$\ $
\includegraphics[width=8cm]{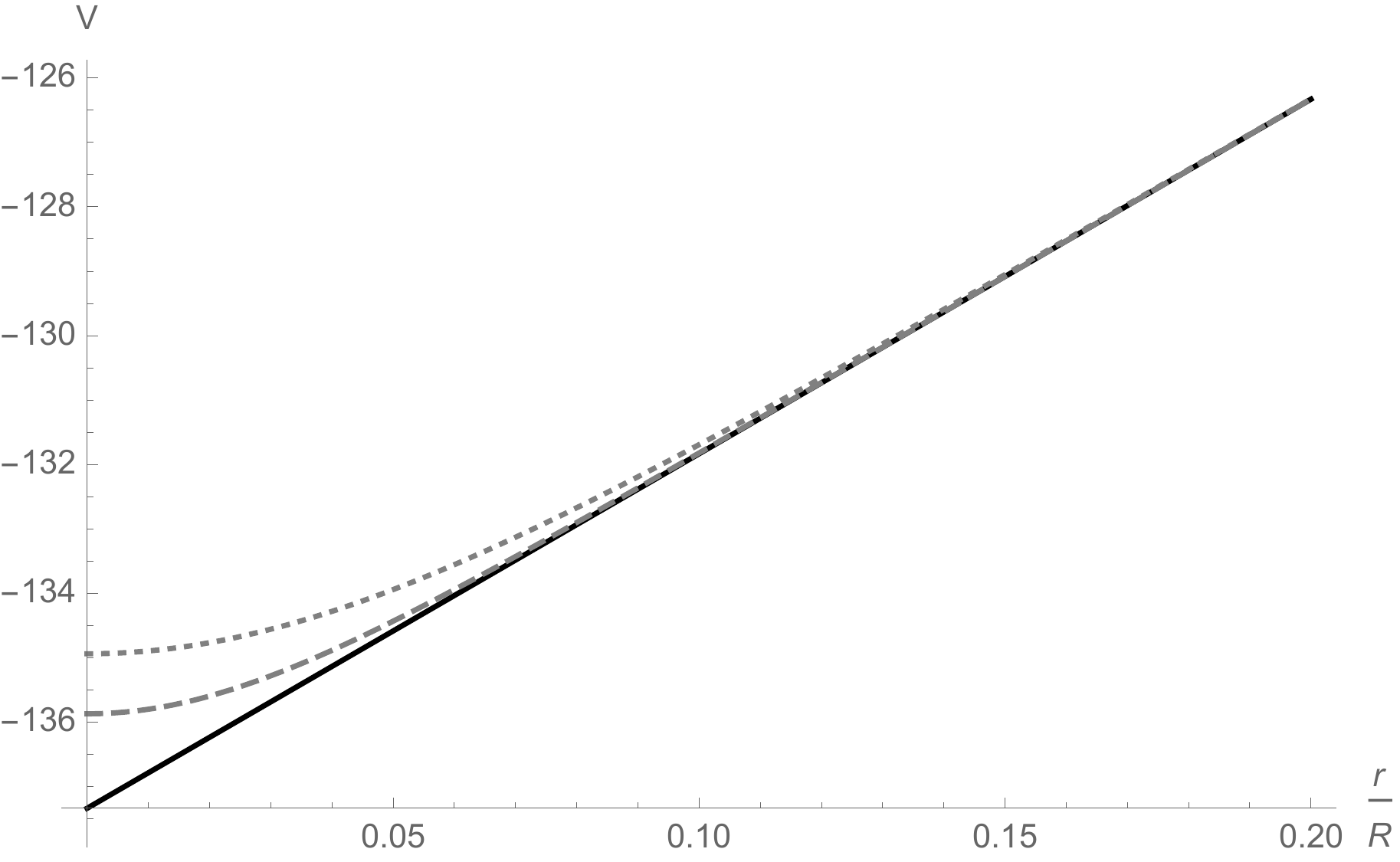}
\caption{Bounding functions $V_-$ (dashed line) and $V_+$ (dotted line) {\em vs\/} linear approximation
(solid line) for $q_\rho=0$ and $x=10^3$.
Bottom panel is a close up view near $r=0$.}
\label{f:1000}
\end{figure}

Bounding functions for Eq.~\eqref{EOMVint} can then be obtained as
\be
\label{VPM}
V_\pm
=
C_\pm\,\psi(r;A_\pm,B_\pm)
\ ,
\ee
where $A_\pm$, $B_\pm$ and $C_\pm$ are constants computed by imposing the boundary
conditions~\eqref{b0}, \eqref{bR} and~\eqref{dbR}.
One first determines a function $V_C=C\,\psi(r;A,B)$ and corresponding mass $M_0$ which satisfy
the three boundary conditions for any constant $C$ and, for fixed values of $R$, $X$ and $q_\rho$,
one can then numerically determine a constant $C_+$ such that $E_+<0$ and a constant $C_-<C_+$
such that $E_->0$.
For example, for the limiting case $q_\rho=0$ and $X=10^3$, we obtain $C_+\simeq 1.73$ and $C_-\simeq 1.05$.
The two bounding functions are then plotted in Fig.~\ref{f:1000} along with the linear approximation~\eqref{Vlin}.
For a comparison, we recall that $C_+\simeq 1.6$ and $C_-\simeq 1$ for $q_\rho=1$ and $X=10^3$ from
Ref.~\cite{Casadio:2019cux}.
\end{document}